# Decoy State Quantum Key Distribution Using Homodyne Detection

S. H. Shams Mousavi[1], and P. Gallion[2]

1. Ecole Supérieure d'Électricité (Supélec), Photonic and Communication Systems, 2 rue Edouard Belin, 57070 Metz, France
2. TELECOM ParisTech, Ecole Nationale Supérieure des Télécommunications, CNRS LTCI UMR 5141, 46 rue Barrault, 75013 Paris, France

In this paper, we propose to use the decoy-state technique to improve the security of the quantum key distribution (QKD) systems based on homodyne detection against the photon number splitting (PNS) attack. The decoy-state technique is a powerful tool that can significantly boost the secure transmission range of the QKD systems. However it has not yet been applied to the systems that use coherent detection. After adapting this theory to the systems based on homodyne detection, we quantify the secure performance and transmission range of the resulting system.

## I. INTRODUCTION

In 1984, Bennett and Brassard first proposed the quantum key distribution to achieve perfect security in optical communication systems [1]. The security in these systems does not have the presumption of the eavesdropper's limited computational power unlike the conventional cryptographic methods and it only relies on the principles of the quantum mechanics and the physics of the channel.

However, a series of attack scenarios can be assumed against QKD systems and need to be addressed, including the individual attacks, the third party attacks and the PNS attack.

Lütkenhaus et al. analyzed the security of the QKD systems against the individual attacks [2]. However, this security analysis is only valid for the QKD systems that use the so-called standard photon counting detection method. Another category of the QKD systems uses the homodyne detection instead of photon counting and requires specific security analysis that was first addressed by Namiki and Hirano [5].

The next step is to consider the third party attacks. These attacks exploit the basis dependant flaws in the source and the detectors of the practical QKD setups. The Gottesman, Lo, Lütkenhaus and Preskill (GLLP) group presented a complete analysis of the QKD system [6] which had considered both the individual attacks and the third party attacks and resulted in a tighter bound on the secure transmission rate compared to the bound found by Lütkenhaus et al.

Finally, the most powerful attack that we can assume against the QKD systems is the PNS attack. The PNS attack exploits the fact that the usual weak coherent state (WCS) sources may send multiple photon pulses instead of single photon pulses that was assumed in the theory. The decoy state technique has been proposed to overcome this specific type of attack [11]. The effect of the PNS attack on the QKD systems using photon counters has already been analyzed and the decoy state technique has been applied to them to improve their security. In this article, we address this problem for the homodyne detection based systems. Our practical setup uses optical homodyne detection which is a special type of the coherent detection allowing a single quadrature measurement of the optical field and therefore a limitation by the zero point fluctuation entering in the signal port. We first measure the impact of the PNS attack on the security of the system. Then, we apply the decoy state technique to improve the secure performance of this QKD system against this attack.

In section II, we first review some of basics of the homodyne detection. Next, we recall the mechanism of the PNS attack and its impact on the security of the system in section III. Then, in section IV, we present the decoy state technique and apply it to our system which is based on homodyne detection. The section V is dedicated to our experimental setup and our measurements. Eventually, we make a conclusion in section VI and give some remarks on our results and the future perspectives.

## II. HOMODYNE DETECTION

Traditionally, the photon counters are most often used in QKD systems because of their capability of operating in low signal levels. We use the homodyne detection instead because of its advantages over the photon counting, most importantly its better quantum efficiency in the optical communication systems wavelength, its ability to deal with the phase encoded signals at very high clock rates, its thermal noise free operation as far as a strong local oscillator is available and its lower cost.

The quantum bit error rate (QBER) in homodyne detection can be obtained from the equation (1) [7].

$$e = 0.5\, erfc\left(\sqrt{2\eta\mu}\right) \quad . \quad (1)$$

where $\eta$ is the channel efficiency and can be calculated for any given channel length, $\ell$, and the attenuation coefficient of the channel, $\alpha$; i.e. $\eta = 10^{-\alpha\ell}$. $\mu$ is the average number of photons per pulse.

For the WCS systems, the typical value of QBER in homodyne detection is higher than the one typically found in the photon counting and it results in a lower performance. However, we can improve this QBER by using a double symmetrical threshold decision discarding weak signals [8], equation (2).

$$e = 0.5\, erfc\left[\sqrt{2\eta\mu}(1+x)\right]. \quad (2)$$

In equation (2), $x$ is the normalized value of the threshold; i.e. we normalize the average level of the receiving signal to one.

The drawback to this change in the decision threshold is that we abandon a portion of the receiving qubits with the following bit rate, bit abandonment rate (BAR).

$$a = 0.5\, erfc\left[\sqrt{2\eta\mu}(1-x)\right] - e \quad . \quad (3)$$

Therefore, in practice, we need to make a compromise between the BER and BAR in order to get the optimum information throughput.

## III. PHOTON NUMBER SPLITTING

In the basic BB84 protocol, Bennett and Brassard assumed a perfect single photon source. However, this type of light source has been out of reach up to now and in practical setups, we use the WCS laser sources with approximately a Poissonian photon number distribution (PND). This special shortcoming of the practical systems makes them vulnerable against the photon number splitting (PNS) attacks.

In order to perform the PNS attack, the eavesdropper needs a non-demolition photon counter [9] to determine the number of photons in each pulse without disturbing its quantum state. Then, the optimum strategy is to block all single photon pulses and to steal one or more photons from the multi-photon pulses in order to get the full information on the key.

This attack clearly adds up some extra loss to the receiving signal. If this added loss is not compensated by Eve, it will change the QBER and therefore the PND at the Bob's end and the attack can be detected by Bob.

Thus, Eve needs to compensate this added loss by replacing the whole or a part of the transmission medium by a superior channel with a lower attenuation. We assume a replacement by a perfect channel with no attenuation, at the worst case. She should also maintain the Poissonian PND.

It has been shown that the PNS attack is detectable, if the following inequality holds true [10].

$$(1+\mu+\frac{\mu^2}{2}).e^{-\mu} - (1+\eta\mu).e^{-\eta\mu} \geq 0. \quad (4)$$

For the numerical values of $\mu = 1.65$ and $\alpha = 0.21$, we find the maximum channel length of only 1.24 km independent of the type of the QKD system and the detection scheme, which is clearly not enough for the practical systems.

But, we should also consider the other types of attacks. In general, the secure transmission rate in any QKD system can be found from the following equation.

$$R = q \cdot p_D \cdot \eta_{pp}, \quad (5)$$

where $q = 0.5$ for the standard BB84 protocol and the detection probability, $p_D$ can be easily calculated for the Poissonian source; i.e. $p_D = 1 - \exp(-\eta\mu)$. It is only the third factor $\eta_{pp}$, called the post processing efficiency, which makes all the difference among the QKD systems.

According to the GLLP security analysis, the secure value of the $\eta_{pp}$ is equal to the maximum of $\eta_1$, obtained from equation (6), and zero [6].

$$\eta_1 = -f(\delta)H(\delta) - (1-\Delta)\left[1 - H(\delta/(1-\Delta))\right], \quad (6)$$

where $\Delta$ is the detection probability of the multi-photon pulses over the overall detection probability ratio, i.e. $\Delta = p_M / p_D$. The parameter $\delta$ is the QBER and the function $f(\delta)$ is the efficiency of the error correcting code used in the post processing; we assume the typical value of $f(\delta) = 1.22$. The function $H(.)$ is the standard binary entropy function.

Figure (1) displays the transmission rate as a function of the channel length for the case of the QKD system based on homodyne detection. As we can see, in this case, the GLLP bound is even more stringent than the bound imposed by the PNS attack.

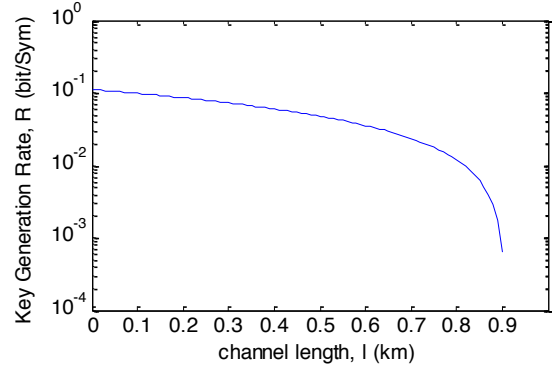

**Figure 1.** GLLP transmission rate (R) for the homodyne detection system

## IV. DECOY-STATE QKD

One powerful method to counteract the threat posed by the PNS attack against the QKD systems is the decoy state technique.

We explain the idea behind the decoy state technique in a very simple way. While performing the PNS attack, the eavesdropper needs to maintain the QBER. Otherwise, the attack will be detected by the legitimate parties. Now, we randomly choose a portion of the qubits and send them with a different coherent state called the decoy state, several decoy states can also be used.

At the end of the transmission, Alice announces publicly for each qubit if she has used the decoy state or the signal state. Then, Bob calculates the QBERs of the signal state and the decoy state. Since the signal and decoy states are randomly chosen and since two coherent states are non-orthogonal, Eve can not distinguish one from each other and at best, she can maintain one QBER. So, Bob can detect the eavesdropping by checking both QBERs providing that they are reasonably apart.

Now, by combining the decoy-state theory with the GLLP security analysis, we can find the secure transmission rate against individual attacks, third party attacks and the PNS attack, all at the same time. We can choose to either use the decoy state for the actual transmission of the data or to simply send dummy bits while using the decoy state. In general, the post processing efficiency for each state is the maximum of zero and $\eta_1$ obtained from the following equation.

$$\eta_1 = (1-2a)\left[-f(e)H(e) + (1-\Delta)[1-H(e)]\right]. \quad (5)$$

The term $1-2a$, in the right hand side of the equation (5), is due to the BAR of the coherent detector with the displaced threshold. For the values of threshold less than one, the transmission rate resulted from equation (5) has the typical form of the figure (2) with a maximum transmission rate at zero channel length and a maximum channel length in which the transmission rate drops down to zero.

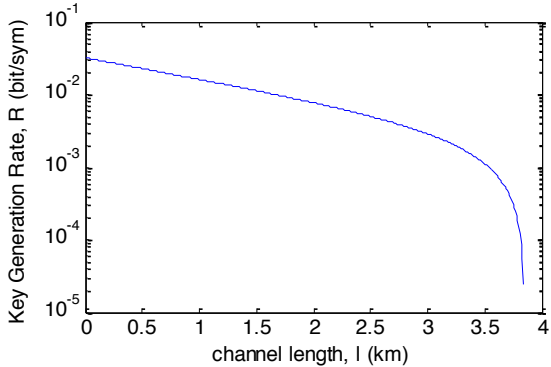

**Figure 2.** Secure transmission rate (R) vs. channel length (l) with $\mu = 1.65$ and x = 0.9.

Interestingly, with the detection thresholds higher than one we also observe a minimum channel length, figure (3). This is due to the fact that while the BER decreases by increasing *x*, the BAR increases. There is obviously an optimum value for the *x* that gives the highest bit rate.

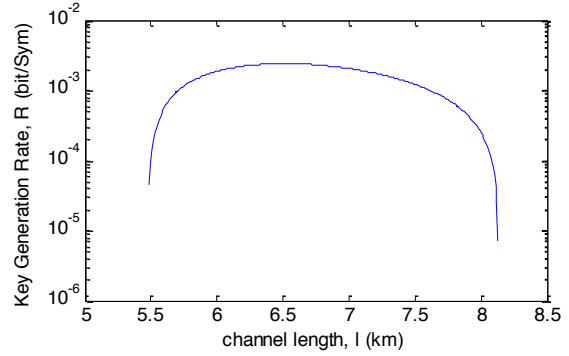

**Figure 3.** Secure transmission rate (R) vs. channel length (l) with $\mu = 1.65$ and x = 2.

Generally, by increasing the threshold, both the minimum and the maximum channel lengths increase (figure 4). For the thresholds lower than one, the maximum R always happens at $\ell = 0$. However, for higher values of the threshold the best performance occurs at a particular channel length, also plotted in figure (4). We have drawn the maximum R for the values of threshold between 0 and 10 in figure (6).

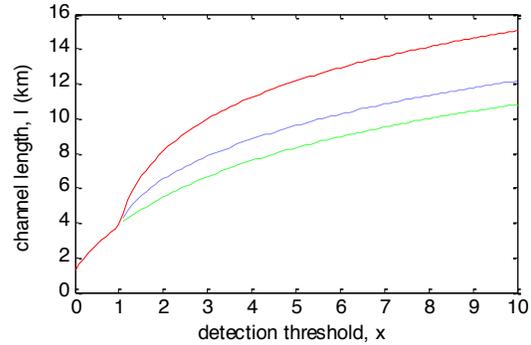

**Figure 4.** Minimum, optimum and maximum channel lengths, respectively from bottom to top for different values of threshold (x)

In summary, the value of the threshold should be chosen according to the variations of the minimum and maximum channel lengths so that the actual channel length will be between these two bounds, with the best case being the channel length that gives the optimal performance. Furthermore, we can always increase the operational range of the system by increasing the threshold at the expense of a decrease in the transmission rate.

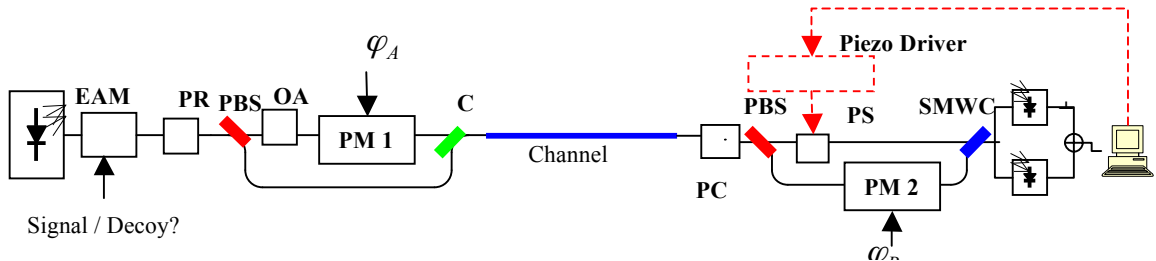

**Figure 5.** Block diagram of the weak coherent pulse quantum key distribution system using balanced homodyne detection. EAM: Electro-Absorbent Modulator, PR: Polarization Rotator, OA: Optical Attenuator, PM: Phase Modulator (Mach-Zehnder type), C: 3 dB coupler, PC: Polarization Controller, PS: Phase Shifter, PBS: Polarization Beam Splitter (combiner), SMWC: Single Mode Wideband Coupler.

Finally, we should add that if we choose to send the information by several coherent states, the overall transmission rate will be the sum of all bit rates each multiplied by the percentage of the qubits sent by the corresponding coherent state; e.g. $R = p_1 R_1 + p_2 R_2$ for two coherent states.

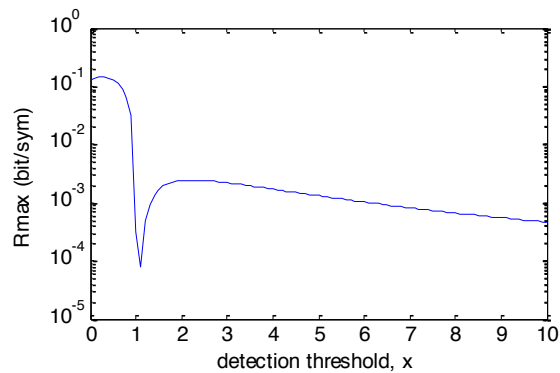

**Figure 6.** Maximum transmission rate (R) for the different values of the threshold (x)

## V. EXPERIMENTAL RESULTS

For the experiments, we have used a QKD system based on homodyne detection that is implemented in our laboratory [8]. We tested the system for the short channel lengths (a few meters) to find its secure performance. Figure (5) shows the block diagram of our system.

We set the value of the threshold at zero and performed a series of experiments for several values of $\mu$. These results and the expected theoretical curve are shown in figure (7). We found the best performance at $\mu = 1.65$, which was used in all simulations along the paper.

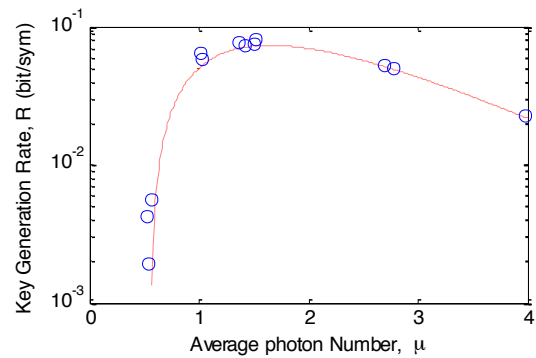

**Figure 7.** Transmission rate (R) vs. $\mu$ for the threshold at zero, x = 0

This setup is an implementation of the standard four-state BB84 protocol via phase encoding. The interferometric structures in two sides are used to produce the reference pulse and the interference between the signal and reference pulses, respectively. At Alice's side, the first interferometer also allows sending both the signal pulse and the reference pulse through a single channel by the means of a time division. This time division is compensated at Bob's side by adjusting the lengths of the two arms of the second interferometer to let the interference happen.

We compensate the phase instability that is introduced in two interferometers by the means of a feedback loop in the receiver.

The choice of the signal or decoy state is done at the EAM which is also used to make digital pulses out of the continuous laser beam. For a more detailed explanation of the setup, we refer the readers to the reference [8].

The value of μ in these experiments has been estimated using the phase photon-number uncertainty equation.

$$\Delta n \cdot \Delta \varphi \geq \frac{1}{2} \quad . \quad (6)$$

The output of the laser source is approximately a coherent state with the Poissonian PND, For which, we have:

$$(\Delta n)^2 = \mu \quad . \quad (7).$$

Now assuming the minimum uncertainty, i.e. the equality in equation (6), we have:

$$\mu = \sqrt{\frac{1}{2\Delta \varphi}} \quad . \quad (8)$$

The parameter $\Delta \varphi$ in equation (8) is the standard deviation of the detected phase and can be expressed in terms of the average value of detector output voltage (V) and its standard deviation ($(\Delta V)^2$) as:

$$\Delta \varphi = \Delta V / V \quad . \quad (9)$$

## VI. CONCLUSION

In this paper, we demonstrated the application of the decoy state technique to the QKD systems based on homodyne detection to achieve the security against the PNS attacks.

We calculated the performance of the system for a wide range of detection thresholds. From the results, we conclude that for low channel lengths (below 4 km), it is better to set the threshold close to zero. On the other hand, for a longer channel, higher threshold values result in a better performance. Eventually, for any given channel length, we can find the optimum value of threshold by a simple simulation.

The experiments that we have performed are only for a very short channel length in which the attenuation is negligible, further experiments have to be done to validate our results for longer channel lengths which need more accurate devices.

## ACKNOWLEDGMENTS

We would like to thank Qing Qu and Manuel Sabban for the valuable discussions. This work has been supported financially by Agence Nationale de la Rechereche (ANR), France as a part of the HQNET project.